\documentclass[amsmath,amssymb,pacs,prd]{revtex4}

\usepackage{graphicx}
\usepackage{dcolumn}
\usepackage{bm}
\usepackage{subfigure}
\usepackage{float}
\begin{document}

\title{Ground states for nonuniform periodic Ising chains}

\author{J.P. Mart\'{\i}nez-Garcilazo}
 \affiliation{Benem\'erita Universidad Aut\'onoma de Puebla, Facultad de Ciencias
F\'{\i}sico Matem\'aticas, P.O. Box 165, 72000 Puebla, M\'exico.}
 
\author{C. Ram\'{\i}rez}%
 \email{cramirez@fcfm.buap.mx}
 \affiliation{Benem\'erita Universidad Aut\'onoma de Puebla, Facultad de Ciencias
F\'{\i}sico Matem\'aticas, P.O. Box 165, 72000 Puebla, M\'exico.}

 \date{\today}

\begin{abstract}
 We give a generalization of Morita's works on ground states of Ising chains, for chains with a periodic structure with different spins, and distant neighbor interactions. The main assumption is translational invariance. The length of the irreducible blocks is a multiple of the period of the chain. In the case of parity invariance, it restricts the length in general only in the diatomic case. There are degenerated states and under certain circumstances there could be nonregular ground states. We illustrate the results and give the ground state diagrams in several cases.
\end{abstract}
\maketitle

\section{Introduction}
The Ising model has been an important means in the study of statistical systems, since its formulation in 1925. It is an arrangement of spins with exchange interaction, located on the sites of a lattice and which for a certain
configuration of the spins has a hamiltonian
\begin{equation} 
H=-\sum_{i<j}J_{ij}S_iS_j-\sum_ih_iS_i,\label{energia}
\end{equation} 
where $J_{ij}$ are the interaction coupling constants and $h_i$ is an external magnetic field. Each spin interacts with its neighbors up to some range and, for chains, in order to avoid doubling, the interactions are accounted from the left to the right, i.e. the energy of a spin is taken to be the one corresponding to its interactions to its right. In general periodic boundary conditions are imposed, although for our purposes they are not necessary. The exchange interactions in (\ref{energia}) could be of a more general form, for example biquadratic.

In an uniform configuration, translational invariance along the lattice requires that $J_{ij}$ depend on the relative position of the lattice sites, i.e. $J_{ij}=J(|i-j|)$, and the magnetic field is constant. These chains are invariant under inversion or parity. The study and classification of classical ground states for Heisenberg chains has been pursued also for a long time \cite{bethe,gosh}. For the Ising chain Luttinger and Tisza conjectured that it should be given by a regular lattice, formed by the repetition of an elementary block, and as quoted by Karl \cite{karl}, the minimal configuration of classical spins is the usual starting point for the quantum treatment. In this paper, \cite{karl}, a proof of the conjecture under certain assumptions has been given, mainly that the spin-spin interactions are parity invariant, and the proof based on the decomposition of the energy expression into several terms which are then minimized.

The ground states for uniform chains with a general form of the exchange energy, have been thoroughly studied by Morita \cite{morita}, who generalized previous work \cite{moritah,katsura} and has shown that the energy of a chain can be written as the sum of energies of irreducible blocks, from which the ground states turn out. The results of Morita can be seen to rely basically on translational invariance. Moreover, there is a certain degree of degeneracy under ground states due to translational invariance \cite{morita}, which puts limits on the maximum length of irreducible blocks, which is further constrained by invariance under parity transformation \cite{bundaru,morita2}.  

In this work we generalize Morita's approach to chains
which are nonuniform at short scale but uniform at large scale, i.e. they are periodic, with site dependent couplings and in general different spins.
They could be used to modelate for example polymers \cite{strobl}, lattice gases \cite{lee} and through competing interactions could arise frustration \cite{sandvik}. 
If the period is $l$, there are $l$ different types of atoms (spins), which interact with the other spins in a translational invariant way, but with a coupling dependent on the atom type or site. Thus a chain of this sort will have the form $\cdots S^{(1)}_i\cdot\cdot S^{(l)}_{i+l}\cdots S^{(1)}_j\cdot\cdot S^{(l)}_{j+l}\cdots$, where the upper index denotes the atom type, and the lower index denotes the position on the chain. Further, as the labeling is a matter of convention, if we speak on molecules, they could begin at any place in the chain, then repeating regularly. Thus when we refer to molecules, we will understand substructures with this ambiguity. 

The exchange energy between two spins is usually given by 
\begin{equation}
J^{(a)}_{|j-i|}\phi(S^{(a)}_i,S^{(b)}_j),\label{interaccion}
\end{equation}
where the dependence of the coupling on the type of the second spin is implicit in the dependence on its position. The function $\phi$ can be as usual bilinear, biquadratic, etc. Morita has considered	 generalized multispin interactions of range $r$, with Hamiltonian
\begin{equation}
H=\sum_{i}\phi(S_i,S_{i+1},\cdots ,S_{i+r}),\label{hg}
\end{equation}
where the function $\phi$ is translational invariant. This form can be generalized for polyatomic chains and we show that the length of irreducible blocks satisfy the same main criteria of Morita. Further, there is an equivalence of an irreducible block and its cyclic permutations and the length of an irreducible block is a multiple of $l$. In general there is parity invariance only when the molecules are parity invariant, otherwise the highest length of irreducible blocks depends on the period, the interaction range and the spin values. Morita sets limits of the maximal length of ground state blocks due to parity invariance \cite{morita2}, we give an explicit demonstration of these results. The ground states are obtained by direct energy comparison in the parameter space of the chain and we give the irreducible blocks in various examples and their ground state diagrams. 
The outline of the paper is the following. In Sec. 2 we review Morita's formulation \cite{morita,morita2}, in Sec. 3 we give the generalization to diatomic chains with distant neighbor interactions, in Sec. 4 the generalization is given for polymeric chains, in Sec. 5  the ground state diagrams for some examples are given and in Sec. 6 conclusions are drawn.

\section{Morita's formulation}
In this section we will consider Morita's formulation for the reduction of a one-dimensional homogeneous chain \cite{morita,morita2} with Hamiltonian given by (\ref{hg}). Morita defines ``displaceable'' blocks and by means of them the reducibility of a chain, which allows to obtain the ground states as irreducible states of minimum energy. Additionally he establishes a maximal length for the irreducible states, from which regular chains are constructed. We are not going into the explicit details of Morita's approach. Instead, we make a formulation in different and somewhat simpler terms, which will be then generalized for multiatomic chains.

Take first the simplest Ising chain ${\cal C}$ with $N$ spins $S=\pm 1$ and next nearest neighbor interaction. In the following we will consider for simplicity bilinear Hamiltonians, with constant magnetic field, although the steps done and the results do not depend on this form and are valid for the hamiltonian (\ref{hg}) 
\begin{equation}
H({\cal C})=-\sum_{i=1}^{N}(J S_{i}S_{i+1}+hS_i),
\end{equation}
where in general the chain is periodic, but if it is finite, then $S_i=0$ for $i>N$ or $i<1$. 

If 2 adjacent spins have the same value $U$, i.e. ${\cal C}=\underbrace{S_{1}\cdots S_n}_{{\cal C}_0}U\underbrace{US_{n+2}\cdots S_{N}}_{{\cal C}_1}$, then the energy of the chain can be written as
\begin{equation}
E({\cal C})=-\sum_{i=1}^{n}(J S_{i}S_{i+1}+hS_i)-(J+hU)-\sum_{i=n+2}^{N}\left(J S_{i}S_{i+1}+hS_i\right),
\end{equation}
where $S_{n+1}=S_{n+2}=U$. Due to translational invariance the middle term on the r.h.s does not depend on its position in the chain. This term is the energy of the spin $U$ in such a way that to its right there is a similar spin. In general, with the convention that the interactions of a spin are accounted to the right, the energy of a block ${\cal B}$ whose interactions are with the $r$ spins of a block ${\cal K}$, where $r$ is the interaction range, will be written as $E({\cal B};{\cal K})$. However we will write simply $E({\cal B})$ when ${\cal K}$ coincides with the first $r$ spins of ${\cal B}$, or as it is explained in the following when the length of ${\cal B}$ is less than $r$.
Thus
\begin{equation}\label{energia1}
E({\cal C})=E({\cal C}_0)+E(U)+E({\cal C}_1).
\end{equation}
We suppose that $N$ is large enough in order that the contribution of the boundary terms is not relevant, although the results are in general valid also for shorter chains, with account of the boundary terms. Therefore a chain of this type, with spin values $S=\pm 1$ and $n_{\pm}$ repeated spins $\pm 1$, will have energy $E({\cal C})=n_{+}\epsilon_{+}+n_{-}\epsilon_{-}+E({\cal \tilde C})$, where $\epsilon_{\pm}=-J\mp h$ are the corresponding energies and the chain ${\cal\tilde  C}$ does not contain repeated spins, i.e. it is formed by blocks of up-down spins $\uparrow\downarrow$. Thus the energy of the chain can be reduced successively to a sum of the energies of the blocks which compose it, i.e. ${\cal C}_{+}=\uparrow$, ${\cal C}_{-}=\downarrow$ and ${\cal C}_{+-}=\uparrow\downarrow$, and 
\begin{equation}
E({\cal C})=n_{+}\epsilon_{+}+n_{-}\epsilon_{-}+2n_{+-}\epsilon_{+-},\label{mono}
\end{equation}
where $n_{+-}$ is the number of blocks $\uparrow\downarrow$ and $\epsilon_{+-}=1/2E({\cal C}_{+-})=J$ are the energies per spin, i.e. the energies divided by the number of spins of the block. Note that the energy of a block $\downarrow\uparrow$ is the same, $\epsilon_{-+}=J$. One of these three energies will be minimum in some region of parameter space, hence the configuration given by the repetition of the corresponding block will be a ground state of this chain. 

For higher spin, with a number of values $\mathcal{S}$, equation (\ref{mono}) will be now
\begin{equation}
E({\cal C})=\sum_{u=1}^{\cal S}n_{u}\epsilon_{u}+\sum_{u>v} n_{uv}E(S_uS_v).\label{monos}
\end{equation}
Thus there are $1/2{\cal S}({\cal S}+1)$ irreducible states with spin energies $\epsilon_u=-J(S_u)^2-hS_u$ and $\epsilon_{uv}=-JS_uS_v-h/2(S_u+S_v)$, $u\neq v$.

Consider now a chain with second nearest neighbor interactions, with first and second neighbor couplings  $J_1$ and  $J_2$. Let ${\cal C}=\underbrace{S_{1}\cdots S_n}_{{\cal C}_0}U\underbrace{UUS_{n+3}\cdots S_{N}}_{{\cal C}_1}$, its energy is also given by (\ref{energia1}), but now $E(U)=-(J_1+J_2+h)$. Let us take in general ${\cal C}=\underbrace{S_{1}\cdots S_n}_{{\cal C}_0}{\cal B}\underbrace{U_1U_2S_{n+k+2}\cdots S_{N}}_{{\cal C}_1}$, where ${\cal B}=U_1U_2\cdots U_k$ and $k>1$. Its energy is 
\begin{equation}\label{energia2}
E({\cal C})=E({\cal C}_0)+E({\cal B})+E({\cal C}_1)=E({\cal B})+E({\cal C}_0{\cal C}_1)
\end{equation}
where ${\cal C}_0{\cal C}_1$ is a chain formed by ${\cal C}_0$ and ${\cal C}_1$ placed one after the other, and
\begin{equation}\label{energiab}
E({\cal B})=-\sum_{i=1}^{k}(J_1 U_{i}U_{i+1}+J_2 U_{i}U_{i+2}+hU_i)
\end{equation}
and $U_{k+1}=U_{1}$, $U_{k+2}=U_{2}$.
We call irreducible a block ${\cal B}$, whose energy is (\ref{energiab}), if the regular chain ${\cal B}\cdots{\cal B}$ cannot be, nontrivially, reduced in the same way as (\ref{energia2}). A trivial reduction would be to reduce it by the same block ${\cal B}$. Thus if a block is irreducible, a cyclic permutation of its spins is also irreducible. Consistently with it, it is easy to see that (\ref{energiab}) is invariant under cyclic permutations of $U_1U_2\cdots U_k$. Hence an irreducible block can be represented by any its cyclic permutations. 

Therefore, for $S=1/2$ and nearest neighbor interactions, the only irreducible blocks are: for one spin $\uparrow$ and $\downarrow$, for two spins $\uparrow\downarrow$, for three spins $\uparrow\uparrow\downarrow$ and $\uparrow\downarrow\downarrow$, and with four spins $\uparrow\downarrow\downarrow\uparrow$.  

Following this construction, it is straightforward to generalize reducibility for a chain ${\cal }$ with $r$-neighbor interactions. The idea is that there is a block ${\cal B}$ in the chain, whose energy $E({\cal B})$ is the same as in a regular chain of this block, i.e. formed by repetition of itself. In this case, the energy of ${\cal C}$ decomposes into the sum of $E({\cal B})$ plus the energy of the rest, and the rest assembles naturally into another chain and could reduce further, as in (\ref{energia2}).

There are two cases, for a block ${\cal B}=U_1\cdots U_k$ with $k<r$ spins, we must have 
\begin{equation}
{\cal C}=\underbrace{S_{1}\cdots S_n}_{{\cal C}_0}{\cal B}\underbrace{\overbrace{{\cal B}\cdots{\cal B}}^lU_1\cdots U_mS_{n+k(l+1)+m+1}\cdots S_{N}}_{{\cal C}_1},\label{c1}
\end{equation}
where $l$ is the greatest integer such that $kl<r$ and $m=r-kl$. 

If $k>r$, then the chain must satisfy 
\begin{equation}
{\cal C}=\underbrace{S_{1}\cdots S_n}_{{\cal C}_0}{\cal B}\underbrace{U_1\cdots U_r S_{n+k+r+1}\cdots S_{N}}_{{\cal C}_1}.\label{c2}
\end{equation}
In both cases the energy is given by (\ref{energia2}). These two cases are resumed by the displaceable blocks of Morita \cite{morita}, which means that there are two identical $r$-lenght blocks ($r$-blocks) at different places in the chain. If these blocks intersect we have (\ref{c1}) and in the contrary case  (\ref{c2}).  Hence the length of a reducible block must be at least $r+1$, i.e. all blocks with length $k\leq r$ are irreducible. Further, if a chain is long enough, there will be necessarily displaceable blocks. Indeed \cite{morita}, if the number of spin values is $s_M$, then a chain of length $s_M^r+r$ will contain at least one displaceable block, because this chain has $s_M^r+1$ succesive $r$-blocks, and there are only $s_M^r$ different such blocks. Thus the length of an irreducible block can be at most $s_M^r+r-1$. However, the equivalence under cyclic permutations imposes a further restriction on these blocks (for $r>1$). Indeed, a maximal length irreducible block ${\cal B}$ can be constructed by arranging the $s_M^r$ different $r$-blocks as in the preceding case. 
Then, cyclic equivalence allows us to set at the beginning of the block, an $r$-block with equal spins $S\cdots S$, hence the next $r$-block will be $S\cdots\ SS'$, with $S'\neq S$. Further, there are $s_M-1$ $r$-blocks $S''S\cdots\ S$ and $s_M-2$ $r$-blocks $S\cdots\ SS'''$, $S'',S'''\neq S$, $S'''\neq S'$. $s_M-2$ of the former blocks will appear together with the last ones, as the $(r+1)$-blocks $S''S\cdots\ SS'''$. Thus the remainig $r$-block $S''S\cdots\ S$ must be the last one, and by cyclicity, its $r-1$ spins $S$ can be located at the beginning of ${\cal B}$, hence there will be a block with $2r-1$ equal spins, which is reducible. Therefore, in order that ${\cal B}$ is irreducible, it cannot contain these $r-1$ last spins, and its length cannot exceed $s_M^r$. 

If the lattice is parity symmetric, as is the case of homogeneous lattices, the energy of an irreducible block can decompose as shown in the following, and the maximal length of irreducible blocks decreases \cite{morita2}. As seen in the previous paragraph, the condition that an irreducible block does not contain more than $s_M^r$ 
$r$-blocks leads to a maximal length of $s_M^r$. Among these $r$-blocks, there may be $r$-blocks with parity symmetry around their middle, i.e. of the form $S_{1}\cdots S_{r/2}S_{r/2}\cdots S_{1}$ for $r$ even, and $S_{1}\cdots S_{(r-1)/2}S_{(r+1)/2}S_{(r-1)/2}\cdots S_{1}$ for $r$ odd, we will call these blocks $p$-blocks. There are  $s_m^{[(r+1)/2]}$ different $p$-blocks, where $[a]$ is the integer part of $a$. Thus, if $l>s_{M}^r-s_{M}^{[(r+1)/2]}$, there will be $p$-blocks \cite{morita2}. Let us suppose that an irreducible block ${\cal B}$ contains $n$ $p$-blocks. Then, by means of a cyclic transformation, we can write ${\cal B}={\cal P}_{1}{\cal C}_{1}\cup{\cal P}_{2}{\cal C}_{2}\cup\cdots\cup{\cal P}_{n}{\cal C}_{n}$, where ${\cal P}_{k}$ are $p$-blocks, joined by the blocks ${\cal C}_{k}$. The union symbols $\cup$ take into account that the $p$-blocks can overlap. If $A$ and $B$ overlap, then $A\cup B$ must be taken in the same way as in set theory, i.e. it contains $A$ and $B$ without duplicating elements, in other words $A\cup B=A(B\setminus A\cap B)=(A\setminus A\cap B)B=(A\setminus A\cap B)(A\cap B)(B\setminus A\cap B)$, where the intersection is the overlapping block and the simple product means to place one block as the continuation of the other. We will take in the following as a prescription $A\cup B=(A\setminus A\cap B)B$, i.e. for $A$ we take the part of it not contained in $B$.
If we call $p({\cal B})$ or $\tilde{\cal B}$ the parity transformation of a block, then $\tilde{\cal P}_{k}={\cal P}_{k}$. Further, we define a new block by insertions of ${\cal I}_{i}={\cal P}_{i}{\cal C}_{i}\cup{\cal P}_{i+1}\tilde{\cal C}_{i}$, as follows
\begin{eqnarray}
{\cal B}'&=&{\cal I}_{1}\cup{\cal P}_{1}{\cal C}_{1}\cup{\cal I}_{2}\cup{\cal P}_{2}{\cal C}_{2}\cdots
\cup{\cal I}_{n}\cup{\cal P}_{n}{\cal C}_{n}\nonumber\\
&=&{\cal P}_{1}{\cal C}_{1}\cup{\cal P}_{2}\tilde{\cal C}_{1}\cup{\cal P}_{1}{\cal C}_{1}\cup{\cal P}_{2}{\cal C}_{2}\cup{\cal P}_{3}\tilde{\cal C}_{2}\cup{\cal P}_{2}{\cal C}_{2}\cdots
\cup{\cal P}_{n}{\cal C}_{n}\cup{\cal P}_{1}\tilde{\cal C}_{n}\cup{\cal P}_{n}{\cal C}_{n},\label{cadenap}
\end{eqnarray}
where ${\cal P}_{n+1}\equiv{\cal P}_{1}$.  In the case that there is overlap, say between ${\cal P}_{i}$ and ${\cal P}_{i+1}$, and ${\cal C}_{i}$ is void, then there will be overlap between ${\cal I}_{i}$ and the following ${\cal P}_{i}$ in the first row of (\ref{cadenap}), given by $p({\cal P}_{i}\cap{\cal P}_{i+1})$, hence in this case 
\begin{equation}
{\cal I}_{i}=({\cal P}_{i}\setminus{\cal P}_{i}\cap{\cal P}_{i+1})({\cal P}_{i+1}\setminus p({\cal P}_{i}\cap{\cal P}_{i+1})).\label{iover}
\end{equation} 
Then ${\cal B}'$ reduces and 
\begin{equation}
E({\cal B}')=E({\cal I}_{1})+E({\cal I}_{2})+\cdots +E({\cal I}_{n})+E({\cal B}),\label{energiai}
\end{equation}
Further, (\ref{cadenap}) leads also to 
\begin{eqnarray}
E({\cal B}')&=&2E({\cal P}_{1}{\cal C}_{1};{\cal P}_{2})+E({\cal P}_{2}\tilde{\cal C}_{1};{\cal P}_{1})+\cdots+2E({\cal P}_{n}{\cal C}_{n};{\cal P}_{1})+E({\cal P}_{i+1}\setminus p({\cal P}_{i}\cap{\cal P}_{i+1};{\cal P}_{i})\nonumber\\
&=&2E({\cal B})+E({\cal P}_{1}\tilde{\cal C}_{n}{\cal P}_{n}\tilde{\cal C}_{n-1}\cdots{\cal P}_{3}\tilde{\cal C}_{2}{\cal P}_{2}\tilde{\cal C}_{1}),
\end{eqnarray}
where in the case of (\ref{iover}), the corresponding terms in the first row will be $2E({\cal P}_{i}\setminus{\cal P}_{i}\cap{\cal P}_{i+1};{\cal P}_{i+1})+E({\cal P}_{i+1}\setminus p({\cal P}_{i}\cap{\cal P}_{i+1});{\cal P}_{1})$. Therefore, after a cyclic permutation in the last term of (\ref{cadenap}), we get $E({\cal B}')=2E({\cal B})+E(\tilde{\cal B})$. Thus, if the interactions are parity symmetric then, $E(\tilde{\cal B})=E({\cal B})$ and $E({\cal B}')=3E({\cal B})$, i.e., taking into account (\ref{energiai}), we get \cite{morita2}
\begin{equation}
E({\cal B})=\frac{1}{2}\sum_{i=1}^n E({\cal I}_{i})\label{ei},
\end{equation}
where the lengths of the blocks, i.e. their numbers of spins, satisfy $l({\cal B})=1/2\sum_{i}l({\cal I}_{i})$. From the definition of ${\cal I}_{i}$ it can be seen that their lengths are even. The sum in (\ref{ei}) can decompose farther, if some of the blocks ${\cal I}_{i}$ are reducible

The decomposition (\ref{ei}) is non trivial if it contains at least two different terms, which requires $n\geq2$. However, if $n=2$ and $\tilde{\cal C}_{2}={\cal C}_{1}$, then (\ref{cadenap}) becomes ${\cal B}'={\cal I}_{1}{\cal I}_{1}{\cal I}_{1}$ and ${\cal B}={\cal I}_{1}$. Hence in this last case the reduction requires $n\geq3$. This remains valid when ${\cal P}_{1}$ and ${\cal P}_{2}$ overlap. 
Thus, the condition for ${\cal B}$ decomposing nontrivially, i.e. (\ref{ei}) contains at least two terms, is $l>s_M^r-s_{M}^{[(r+1)/2]}+2$. 

The main assumption in the previous analysis is, up to boundary terms, that the lattice is translational invariant. Moreover the demonstration is facilitated by the convention that the exchange energy of a spin accounts its interactions with the spins to its right (it could be the other way around). Additionally, the parity invariance of the lattice reduces the maximal length of the elementary ground state blocks. 

Therefore, the energy of a chain is the sum of the energies of irreducible blocks. A block ${\cal B}$ is irreducible if its regular chain, ${\cal B}\cdots{\cal B}$ cannot be nontrivially reduced. Hence the cyclic permutations of an irreducible block are equivalent. In this way all irreducible blocks can be obtained. Further, if an irreducible block contains at least 2 or 3 $p$-blocks (see above), its energy decomposes as in (\ref{ei}). If any of the blocks in this decomposition is reducible, it must be reduced and (\ref{ei}) will be expressed in terms of irreducible blocks, whose maximum length is $s_M^r-s_{M}^{[(r+1)/2]}+2$. Therefore, after a full reduction, considering (\ref{ei}), the energy of the chain is, up to boundary terms
\begin{equation}
E({\cal C})=\sum_i n_i E({\cal B}_i)=\sum_i\nu_i\, l({\cal B}_{i})\epsilon({\cal B}_i)\label{ei2},
\end{equation}
where $\nu_{i}=m_{i}+n_{i}/2$, $m_{i}$ are integers and $n_{i}$ are even integers corresponding to the decomposition (\ref{ei}) and $\epsilon({\cal B}_{i})$ are the energies per spin. Therefore, consistently, $l({\cal C})=\sum_i\nu_i\, l({\cal B}_{i})$. Therefore, if ${\cal B}_{0}$ is the minimum energy block in (\ref{ei2}), then the corresponding regular chain will be the ground state.

Moreover, it may be that a block composed by the same irreducible block is as well irreducible, e.g. $\uparrow$ and $\uparrow\uparrow$ for second neighbor interactions. They have the same energy per spin and represent the same ground state. Thus it is enough to consider only the simplest one. 

\section{Diatomic chain}
We generalize the preceding procedure for chains with period 2, with two types of spins $S_i^{(1)}$ and $S_i^{(2)}$, which in general differ by their interactions and their values. The chain has in general the form $S_{1}^{(1)}S_{2}^{(2)}\cdots S_{2k-1}^{(1)}S_{2k}^{(2)}\cdots S_{2N-1}^{(1)}S_{2N}^{(2)}$, although, as mentioned in the first section, it could begin as well with a spin of type $(2)$, differing only by boundary terms. The generalization of (\ref{hg}) is
\begin{equation}
H=\sum_i\left[ \phi_1\left(S_i^{(1)},S_{i+1}^{(2)},\cdots,S_{i+r}^{(1+r)}\right)+\phi_2\left(S_i^{(2)},S_{i+1}^{(1)},\cdots,S_{i+r}^{(2+r)}\right)\right],\label{hdia}
\end{equation}
where $S^{(a)}\equiv S^{(1)}$ if $a$ is odd and $S^{(a)}\equiv S^{(2)}$ if $a$ is even. In the following, as in the previous section, for simplicity we will consider bilinear interactions, although the computations remain valid for (\ref{hdia}). Thus, for first neighbor interactions, the Hamiltonian in the presence of a magnetic field is given by
\begin{equation}
H=-\sum_i\left(J^{(1)}S_{i}^{(1)}S_{i+1}^{(2)}+J^{(2)}S_{i+1}^{(2)}S_{i+2}^{(1)}\right)-h\sum_i\left(S_i^{(1)}+S_i^{(2)}\right),\label{h21}
\end{equation}
where in general the chain will have periodic boundary conditions, but if it is finite, then $S^{(a)}_i=0$ for $i>2N$ or $i<1$. One spin can reduce if it repeats, and it can repeat only after an integer number of periods. Thus reducibility must be defined on blocks of the size $2k$.
Hence, we take
\begin{equation}
{\cal C}=\underbrace{S_{1}^{(1)}S_{2}^{(2)}\cdots S_{2n-1}^{(1)}S_{2n}^{(2)}}_{{\cal C}_0}U^{(1)}U^{(2)}\underbrace{U^{(1)}S_{2(n+2)}^{(2)}\cdots S_{2N-1}^{(1)}S_{2N}^{(2)}}_{{\cal C}_1}. \label{cadena21}
\end{equation}
The energy of this chain is
\begin{eqnarray}
E({\cal C})=&-&\sum_{i=1}^{2n-1}\left(J^{(1)}S_{i}^{(1)}S_{i+1}^{(2)}+J^{(2)}S_{i+1}^{(2)}S_{i+2}^{(1)}\right)-h\sum_{i=0}^{2n-1}\left(S_i^{(1)}+S_{i+1}^{(2)}\right)
-\left(J^{(1)}+J^{(2)}\right)U^{(1)}U^{(2)}-h\left(U^{(1)}+U^{(2)}\right)\nonumber\\
&-&\sum_{i=2(n+k)+1}^{2N-1}\left(J^{(1)}S_{i}^{(1)}S_{i+1}^{(2)}+J^{(2)}S_{i+1}^{(2)}S_{i+2}^{(1)}\right)
-h\sum_{i=2(n+k)+1}^{2N-1}\left(S_i^{(1)}+S_{i+1}^{(2)}\right),\label{energiad}
\end{eqnarray}
where $S_{2n+1}^{(1)}=S_{2(n+1)+1}^{(1)}=U^{(1)}$. Therefore, similarly as for (\ref{energia2}), this energy can be written as
\begin{equation}\label{energiab1}
E({\cal C})=E({\cal C}_0\cup {\cal C}_1)+E(U^{(1)}U^{(2)}),
\end{equation}
where now $E(U^{(1)}U^{(2)})=-\left(J^{(1)}+J^{(2)}\right)U^{(1)}U^{(2)}-h\left(U^{(1)}+U^{(2)}\right)$. Instead of (\ref{cadena21}), we could have ${\cal C}=\cdots S_{2n-1}^{(2)}S_{2n}^{(1)}U^{(2)}U^{(1)}U^{(2)}S_{2(n+2)}^{(1)}\cdots$, with the same result. Thus, following these lines, a block which reduces a diatomic chain has an even number of spins, ${\cal B}=U_{1}^{(1)}U_{2}^{(2)}\cdots U_{2k-1}^{(1)}U_{2k}^{(2)}$, 
\begin{equation}
{\cal C}=\underbrace{S_{1}^{(1)}S_{2}^{(2)}\cdots S_{2n-1}^{(1)}S_{2n}^{(2)}}_{{\cal C}_0}{\cal B}\underbrace{U_1^{(1)}S_{2(n+k+2)}^{(2)}\cdots S_{2N-1}^{(1)}S_{2N}^{(2)}}_{{\cal C}_1},\label{cadena2n1}
\end{equation}
with energy given by (\ref{energia2}). Thus, ${\cal B}$ is irreducible if it cannot be reduced as (\ref{cadena2n1}). As the property of irreducibility is given with respect to the formation of regular chains, if ${\cal B}$ is irreducible, so will be any cyclic permutation of its spins. Further, the energy of all these permutations is the same, as the corresponding regular chain does not change (up to boundary terms).

Under a parity transformation a chain transforms as
${\cal C}\rightarrow\tilde{\cal C}=S_{2N}^{(2)}S_{2N-1}^{(1)}\cdots S_{2}^{(2)}S_{1}^{(1)}$, and its energy $E(\tilde{\cal C})$ is given by $E({\cal C})$, with the couplings interchanged $J^{(1)}\leftrightarrow J^{(2)}$. Thus, the chain is parity symmetric only if it is invariant under $J^{(1)}\leftrightarrow J^{(2)}$, which amounts to $J^{(1)}=J^{(2)}$, as can be seen from (\ref{h21}). In this case the spin types must be different, otherwise the chain would be monoatomic. In this case, if $r$ is odd and greater than 1,  a chain can contain $p$-blocks and it can be seen that an irreducible block will decompose as in (\ref{ei}).

Next we give some examples, for $S^{(1)},S^{(2)}=1/2$, for $k=1$, the irreducible blocks are ${\cal C}_{++}=\uparrow\uparrow$, ${\cal C}_{+-}=\uparrow\downarrow$  and ${\cal C}_{--}=\downarrow\downarrow$ with energies per spin $\epsilon_{++}=-1/2(J^{(1)}+J^{(2)})-h$, $\epsilon_{+-}=1/2(J^{(1)}+J^{(2)})$, $\epsilon_{--}=-1/2(J^{(1)}+J^{(2)})+h$. Note that these energies are symmetric under $J^{(1)}\leftrightarrow J^{(2)}$. For $k=2$, the only block is ${\cal C}_{++--}=\uparrow\uparrow\downarrow\downarrow$, whose energies are $\epsilon^{(1)}_{++--}=1/2(-J^{(1)}+J^{(2)})$ and $\epsilon^{(2)}_{++--}=1/2(J^{(1)}-J^{(2)})$. For $k\geq3$, it is easy to see that there are not irreducible blocks. 

If the spins of both atoms differ, at least the type of the first spin of the block must be specified. For example if $S^{(1)}=1/2$ and $S^{(2)}=1$, where now we give to the spin $S^{(1)}$ the values $\pm 1/2$ and to $S^{(2)}$ the values $\pm1,0$. The irreducible blocks will be $\stackrel{(1)}{\uparrow}\stackrel{(2)}{\uparrow}$ with energy $-1/2(J^{(1)}+J^{(2)})-3/2\,h$, $\stackrel{(1)}{\uparrow}\stackrel{(2)}{\downarrow}$ with energy $1/2(J^{(1)}+J^{(2)})+1/2\,h$, $\stackrel{(1)}{\downarrow}\stackrel{(2)}{\uparrow}$ with energy $1/2(J^{(1)}+J^{(2)})-1/2\,h$,  $\stackrel{(1)}{\uparrow}\stackrel{(2)}{\uparrow}\stackrel{(1)}{\downarrow}\stackrel{(2)}{\downarrow}$ with energy $-J^{(1)}+J^{(2)}$, $\stackrel{(2)}{\uparrow}\stackrel{(1)}{\uparrow}\stackrel{(2)}{\downarrow}\stackrel{(1)}{\downarrow}$ with energy $J^{(1)}-J^{(2)}$, $\stackrel{(1)}{\uparrow}\stackrel{(2)}{\uparrow}\stackrel{(1)}{\downarrow}\stackrel{(2)}{\rightarrow}$ with energy $1/2(-J^{(1)}+J^{(2)})-h$, $\stackrel{(2)}{\uparrow}\stackrel{(1)}{\uparrow}\stackrel{(2)}{\rightarrow}\stackrel{(1)}{\downarrow}$ with energy $1/2(J^{(1)}-J^{(2)})-h$, $\stackrel{(1)}{\downarrow}\stackrel{(2)}{\downarrow}\stackrel{(1)}{\uparrow}\stackrel{(2)}{\rightarrow}$ with energy $1/2(-J^{(1)}+J^{(2)})+h$ and $\stackrel{(2)}{\downarrow}\stackrel{(1)}{\downarrow}\stackrel{(2)}{\rightarrow}\stackrel{(1)}{\uparrow}$ with energy $1/2(J^{(1)}-J^{(2)})+h$.

For second neighbor interactions, the Hamiltonian is
\begin{equation}
E=-\sum_{i=1}^{2N-1}\left(J^{(1)}_1S_{i}^{(1)}S_{i+1}^{(2)}+J^{(2)}_1S_{i+1}^{(2)}S_{i+2}^{(1)}+J^{(1)}_2S_{i}^{(1)}S_{i+2}^{(1)}+J^{(2)}_2S_{i+1}^{(2)}S_{i+3}^{(2)}\right)-h\sum_{i=0}^{2N-1}\left(S_i^{(1)}+S_{i+1}^{(2)}\right).
\end{equation}
In this case if ${\cal B}=U_1U_2\cdots U_{2k}$ and
\begin{equation}
{\cal C}=\underbrace{S_{1}^{(1)}S_{2}^{(2)}\cdots S_{2n-1}^{(1)}S_{2n}^{(2)}}_{{\cal C}_0}{\cal B}\underbrace{U_1U_2S_{2(n+k)+3}^{(1)}\cdots S_{2N-1}^{(1)}S_{2N}^{(2)}}_{{\cal C}_1}, \label{cadena2n}
\end{equation}
Then the energy is given by (\ref{energia2}).
For $k=1$, with first spin of ${\cal B}$ of type $(1)$, $E({\cal B})=-[(J^{(1)}_1+J^{(2)}_1)U_1U_2+J^{(1)}_2(U_1)^2+J^{(2)}_2(U_2)^2+h(U_1+U_2)]$. In general this energy is not symmetric under $(1)\leftrightarrow(2)$, although there are cases like $S^{(1)},S^{(2)}=\pm 1$ for which it is symmetric. In this last case the irreducible blocks are ${\cal C}_{++}=\uparrow\uparrow$, ${\cal C}_{+-}=\uparrow\downarrow$ and ${\cal C}_{--}=\downarrow\downarrow$ with spin energies $\epsilon_{++}=-1/2(J^{(1)}_1+J^{(2)}_1+J^{(1)}_2+J^{(2)}_2)- h$, $\epsilon_{+-}=1/2(J^{(1)}_1+J^{(2)}_1-J^{(1)}_2-J^{(2)}_2)$ and $\epsilon_{--}=-1/2(J^{(1)}_1+J^{(2)}_1+J^{(1)}_2+J^{(2)}_2)+ h$. 
For $k=2$, ${\cal B}=U_1U_2U_3U_4$, they are irreducible if $U_1U_2\neq U_3U_4$, or what is the same $U_2U_3\neq U_4U_1$. Thus the irreducible blocks and their spin energies are $\stackrel{\uparrow}{_{(1)}}\stackrel{\uparrow}{_{(2)}}\stackrel{\uparrow}{_{(1)}}\stackrel{\downarrow}{_{(2)}}$, $1/2(-J^{(1)}_2+J^{(2)}_2)- h/2$; $\stackrel{\uparrow}{_{(2)}}\stackrel{\uparrow}{_{(1)}}\stackrel{\uparrow}{_{(2)}}\stackrel{\downarrow}{_{(1)}}$, $1/2(J^{(1)}_2-J^{(2)}_2)-h/2$; $\stackrel{\uparrow}{_{(1)}}\stackrel{\uparrow}{_{(2)}}\stackrel{\downarrow}{_{(1)}}\stackrel{\downarrow}{_{(2)}}$, $1/2(-J^{(1)}_1+J^{(2)}_1+J^{(1)}_2+J^{(2)}_2)$; $\stackrel{\uparrow}{_{(2)}}\stackrel{\uparrow}{_{(1)}}\stackrel{\downarrow}{_{(2)}}\stackrel{\downarrow}{_{(1)}}$, $1/2(J^{(1)}_1-J^{(2)}_1+J^{(1)}_2+J^{(2)}_2)$; $\stackrel{\uparrow}{_{(1)}}\stackrel{\downarrow}{_{(2)}}\stackrel{\downarrow}{_{(1)}}\stackrel{\downarrow}{_{(2)}}$,
$1/2(-J^{(1)}_2+J^{(2)}_2)+ h/2$ and $\stackrel{\uparrow}{_{(2)}}\stackrel{\downarrow}{_{(1)}}\stackrel{\downarrow}{_{(2)}}\stackrel{\downarrow}{_{(1)}}$, $1/2(J^{(1)}_2-J^{(2)}_2)+ h/2$. For $k=3$, ${\cal B}=U_1U_2U_3U_4U_5U_6$, which are irreducible if the pairs $U_1U_2$, $U_3U_4$ and $U_5U_6$ are all three different; further, considering the invariance under cyclic permutations, the pairs $U_2U_3$, $U_4U_5$ and $U_6U_1$ must be also different. It turns out that in this case the irreducible blocks are degenerated under the interchange $(1)\leftrightarrow(2)$ and it is not neccesary to specify the spin type. Thus we get 
$\uparrow\uparrow\downarrow\uparrow\uparrow\downarrow$ with energy per spin $1/6(J^{(1)}_1+J^{(2)}_1+J^{(1)}_2+J^{(2)}_2-2h)$ and $\uparrow\uparrow\uparrow\downarrow\downarrow\downarrow$ with energy per spin $1/6(-J^{(1)}_1-J^{(2)}_1+J^{(1)}_2+J^{(2)}_2)$. Further, for $k=4$, there are only two blocks, $\uparrow\uparrow\downarrow\uparrow\uparrow\downarrow\downarrow\downarrow$ and 
$\uparrow\uparrow\uparrow\downarrow\downarrow\uparrow\downarrow\downarrow$, which are degenerated, and whose spin energies are $1/2J^{(1)}_2$ if the first spin is of type $(1)$ and $1/2J^{(2)}_2$ if the first spin is of type $(2)$. There are not irreducible blocks with a higher number of spins. 

For interaction range $r$, the Hamiltonian of the chain is 
\begin{eqnarray}
H&=&-\sum_{i=1}^{N}\Bigg[\sum_{p\geq0}J^{(1)}_{2p+1}S_{i}^{(1)}S_{i+2p+1}^{(2)}+\sum_{p\geq1}J^{(1)}_{2p}S_{i}^{(1)}S_{i+2p}^{(1)}\nonumber\\
&+&\sum_{p\geq0}J^{(2)}_{2p+1}S_{i}^{(2)}S_{i+2p+1}^{(1)}+\sum_{p\geq1}J^{(2)}_{2p}S_{i}^{(2)}S_{i+2p}^{(2)}
+h\left(S_i^{(1)}+S_{i+1}^{(2)}\right)\Bigg].\label{energiadr}
\end{eqnarray}
Let us consider a block ${\cal B}=U_1\cdots U_{2k}$, for which we take the first spin to be of type $(1)$ for definiteness. Thus, as in the previous section, there are two cases: a) if $2k<r$ and b) $2k\geq r$. In the first case we have 
\begin{equation}
{\cal C}=\underbrace{S_{1}^{(1)}S_{2}^{(2)}\cdots S_{2n-1}^{(1)}S_{2n}^{(2)}}_{{\cal C}_0}{\cal B}\underbrace{\overbrace{{\cal B}\cdots{\cal B}}^sU_1\cdots U_tS_{2(n+k)+r+1}^{(a)}\cdots S_{2N-1}^{(1)}S_{2N}^{(2)}}_{{\cal C}_1}, \label{cadena2r1}
\end{equation}
where $s$ is the smallest integer such that $2ks<r$ and $t=r-2ks$. Further $a=1$ if $r$ is even and $a=2$ if $r$ is odd. In the second case, $2k\geq r$, we take 
\begin{equation}
{\cal C}=\underbrace{S_{1}^{(1)}S_{2}^{(2)}\cdots S_{2n-1}^{(1)}S_{2n}^{(2)}}_{{\cal C}_0}{\cal B}\underbrace{U_1\cdots U_rS_{2(n+k)+r+1}^{(a)}\cdots S_{2N-1}^{(1)}S_{2N}^{(2)}}_{{\cal C}_1}. \label{cadena2r2}
\end{equation}
In both cases the energy is given by (\ref{energia2}). Therefore a block ${\cal B}$ is irreducible if the chain with configuration ${\cal B}\cdots {\cal B}$ cannot be reduced by another block as in (\ref{cadena2r1}) or (\ref{cadena2r2}). In order that a block can reduce, its length must be at least $r+2$ and the notion of displaceable block of morita generalizes and requires that there are two identical blocks of length $r$, but now the distance between them must be $2n$. Further, the cyclic permutations of an irreducible block are also irreducible and have the same energy. 

Thus, an $r$-block beginning with any of both types of spin, will be displaceable if there is an identical $r$-block in another position, hence the first spins of both blocks will be at an even distance. Thus, if $s_M^{(1)}$ and $s_M^{(2)}$ are the number of values which can take both types of spins, the maximum number of different possible configurations will be ${\rm C}_M(r)=\left(s_M^{(1)}s_M^{(2)}\right)^{\frac{r}{2}}$ for $r$ even and ${\rm C}_M(r)=\left(s_M^{(1)}s_M^{(2)}\right)^{\frac{r-1}{2}}{\rm max}(s_M^{(1)},s_M^{(2)})$ for $r$ odd. Thus a block will be always reducible if its length is $2{\rm C}_M(r)+r-1$, because it will have the maximum number of possible configurations plus one. 
In the parity symmetric case, with $J^{(1)}=J^{(2)}$, and different spin types $S^{(a)}$ in order that it does not reduce to the monoatomic case, there will be $p$-blocks if $r$ is odd ($r>1$), and the analysis of the previous section can be repeated almost without changes. Indeed, the blocks ${\cal I}$ have even length, hence begin and end at the opposite type of spin, and (\ref{cadenap}) can be implemented.

As an example, for $r=3$, for equal spins $S^{(1)},S^{(2)}=\pm1$ (different couplings), the irreducible blocks are $\uparrow\uparrow$, $\uparrow\downarrow$, $\uparrow\uparrow\uparrow\downarrow$, $\uparrow\uparrow\downarrow\downarrow$, $\uparrow\uparrow\downarrow\uparrow\uparrow\downarrow$, $\uparrow\uparrow\uparrow\downarrow\downarrow\downarrow$, $\uparrow\uparrow\downarrow\downarrow\uparrow\downarrow$, 
$\uparrow\uparrow\uparrow\downarrow\downarrow\uparrow\downarrow\downarrow$, $\uparrow\uparrow\uparrow\downarrow\uparrow\downarrow\downarrow\downarrow$, $\uparrow\uparrow\downarrow\uparrow\downarrow\downarrow\uparrow\downarrow$, plus the ones obtained by the interchange $\uparrow\leftrightarrow\downarrow$, up to parity invariance. 

\section{Polyatomic chain}
The previous procedure can be straightforwardly generalized for polyatomic chains, with $l$ different types of atoms, which different spins and couplings. Such a chain has the form ${\cal C}=S_1^{(1)}\cdots S_l^{(l)}S_{l+1}^{(1)}\cdots S_{2l}^{(l)}\cdots S_{2N-l+1}^{(1)}\cdots S_{2N}^{(l)}$, and there are $l$ different couplings at each level of interaction range. The Hamiltonian (\ref{hg}) is now
\begin{equation}
H=\sum_a\sum_i \phi_a\left(S_i^{(a)}\cdots S_{i+r}^{(a+r)}\right),
\end{equation}
where $S_i^{(a+jr)}\equiv S_i^{(a)}$ for integer $j>1$ and $1\leq a\leq l$. As in the preceding sections, for simplicity we consider bilinear interactions.  
For next nearest neighbor interactions the Hamiltonian is given by
\begin{equation}
H=-\sum_{j=0}^{N-1}\sum_{a=1}^{l}\left(J^{(a)}S_{a+jl}^{(a)}S_{a+jl+1}^{(a+1)}+hS_{a+jl}^{(a)}\right).\label{energiap1}
\end{equation}
Thus, a reducible chain has the form
\begin{equation}
{\cal C}=\underbrace{S_1^{(1)}\cdots S_l^{(l)}\cdots S_{nl+1}^{(1)}\cdots S_{m}^{(a)}}_{{\cal C}_0}{\cal B}\underbrace{U_1S_{2(n+k+1)}^{(2)}\cdots S_{2(n+k+l)}^{(l)}\cdots S_{2N-l+1}^{(1)}\cdots S_{2N}^{(l)}}_{{\cal C}_1}, \label{cadenalr1}
\end{equation}
where 
\begin{equation}
{\cal B}=U_1\cdots U_{kl}\equiv S_1^{(s(a+1))}\cdots S_{kl}^{(s(a+kl))}\label{b}
\end{equation}
Thus the chains ${\cal C}_0$ and ${\cal C}_1$ can be merged and the energy of ${\cal C}$ is given by (\ref{energia2}), where
\begin{equation}
E({\cal B})=-\sum_{a=1}^{l}\sum_{i=0}^{k-1}\left(J^{(a)}U_{a+il}\,U_{a+il+1}+hU_{a+il}\right),\label{energiabp1}
\end{equation}
where $U_{kl+1}\equiv U_1$. Thus irreducibility is defined in the same way as before and the cyclic permutations of an irreducible block are irreducible and all of them have the same energy. For $l\geq 2$, in general there is no parity symmetry because from the right the order of the spins changes and it cannot be restored by a cyclic permutation. It can be seen that there is no symmetry even if parity affects only whole molecules, i.e. if the order inside molecules is not changed.

If all the spins take the same values, then an irreducible block can be specified as ${\cal B}=U_1\cdots U_{kl}$, although its energy  (\ref{energiabp1}) will depend on the type of the spin $U_1$. Thus in general for this block there will be $l$ energies, which can be obtained by the cyclic permutations of the couplings $\{J^{(1)}\cdots J^{(l)}\}$.
For instance for a triatomic chain, with spin $1/2$ for the three atoms, the irreducible blocks with $k=1$ are ${\cal C}_{+++}=\uparrow\uparrow\uparrow$ with energy per spin $\epsilon_{+++}=-1/3(J^{(1)}+J^{(2)}+J^{(3)})-h$, ${\cal C}_{++-}=\uparrow\uparrow\downarrow$ with energies $\epsilon_{++-}^{(1)}=1/3(-J^{(1)}-J^{(2)}+J^{(3)}-h)$, $\epsilon_{++-}^{(2)}=1/3(-J^{(2)}-J^{(3)}+J^{(1)}-h)$ and $\epsilon_{++-}^{(3)}=1/3(-J^{(3)}-J^{(1)}+J^{(2)}-h)$. For $k=2$ we get ${\cal C}_{+++-+-}=\uparrow\uparrow\uparrow\downarrow\uparrow\downarrow$ with energies $\epsilon_{+++-+-}^{(1)}=1/3(J^{(3)}-h)$, $\epsilon_{+++-+-}^{(2)}=1/3(J^{(1)}-h)$ and $\epsilon_{+++-+-}^{(3)}=1/3(J^{(2)}-h)$ and ${\cal C}_{+++---}=\uparrow\uparrow\uparrow\downarrow\downarrow\downarrow$ with energies $\epsilon_{+++---}^{(1)}=1/3(-J^{(1)}-J^{(2)}+J^{(3)})$, $\epsilon_{+++---}^{(2)}=1/3(-J^{(2)}-J^{(3)}+J^{(1)})$ and $\epsilon_{+++---}^{(3)}=1/3(-J^{(3)}-J^{(1)}+J^{(2)})$, and so on for higher $k$.

For range $r$ interactions the Hamiltonian is 
\begin{equation}
H=-\sum_{j=0}^{N-1}\sum_{a=1}^{l}\left(\sum_{s=1}^{r}J^{(a)}_sS_{a+jl}^{(a)}S_{a+jl+s}^{(a+s)}+hS_{a+jl}^{(a)}\right),\label{energiapr}
\end{equation}
where $S_i^{(a)}$ for $a>l$ is defined as previously.

For the generalization of irreducibility, consider again ${\cal B}$ given by (\ref{b}). As in the previous sections, there are two cases: a) $lk<r$ and b) $lk\geq r$. In the first case take 
\begin{equation}
{\cal C}=\underbrace{S_{1}^{(1)}\cdot\cdot S_{l}^{(l)}\cdots S_{l(n-1)+1}^{(1)}\cdot\cdot S_{ln}^{(l)}}_{{\cal C}_0}{\cal B}\underbrace{\overbrace{{\cal B}\cdots{\cal B}}^sU_1\cdots U_tS_{2(n+k)+r+1}^{(a)}\cdots S_{l(N-1)+1}^{(1)}\cdot\cdot S_{lN}^{(l)}}_{{\cal C}_1}, \label{cadenalr1}
\end{equation}
where $s$ is the smallest integer such that $lks<r$, $t=r-lks$ and $a=r'+1$, where $r'<l$ is such that $r=pl+r'$, $p\geq0$. For the second case 
\begin{equation}
{\cal C}=\underbrace{S_{1}^{(1)}S_{2}^{(2)}\cdots S_{2n-1}^{(1)}S_{2n}^{(2)}}_{{\cal C}_0}{\cal B}\underbrace{U_1\cdots U_rS_{2(n+k)+r+1}^{(a)}\cdots S_{l(N-1)+1}^{(1)}\cdot\cdot S_{lN}^{(l)}}_{{\cal C}_1}. \label{cadenalr2}
\end{equation}
In both cases the energy is given by (\ref{energia2}), where $E({\cal B})$ can be calculated by means of (\ref{energiapr}). Therefore a block ${\cal B}$ is irreducible if the regular chain, with configuration ${\cal B}\cdots {\cal B}$, cannot be reduced as in (\ref{cadenalr1}) or (\ref{cadenalr2}). Moreover the cyclic permutations of ${\cal B}$ are also irreducible and all of them have the same energy. 

Thus, in order that a block be displaceable, its length must be $r$ and there must be another identical $r$-block at another place, i.e. the distance of their first spins must be a multiple of the the period $l$. Thus the length of a reducible block must be at least $l+r$. Further, if the number of values of the spins of the chain are $s_M^{(a)}$, then the number of different configurations which can take an $r$-block will be $C_{M}=\max_{a=1}^{l}(s_M^{(a)}\cdots s_M^{(a+r-1)})$. Hence the maximum length of an irreducible block will be $kl$, where $k$ is the maximum integer with $kl\leq l(C_{M}-1)+r$. Similar to the diatomic case, if the molecules are parity symmetric there will be $p$-blocks, which will be centered on a molecule, but in general (\ref{cadenap}) will not be possible, unless the blocks ${\cal I}_i$ have a length multiple of $l$, such that they can be inserted.

For the previous triatomic chain, with $r=2$, the maximum length for irreducible blocks is 9 and, additionally to the irreducible states for $r=1$, for example the state $\uparrow\uparrow\uparrow\uparrow\downarrow\downarrow$ is irreducible, with energies $2(-J^{(2)}_1+J^{(3)}_2-h)$, $2(-J^{(3)}_1+J^{(1)}_2-h)$ and $2(-J^{(1)}_1+J^{(2)}_2-h)$ and $\uparrow\uparrow\uparrow\uparrow\downarrow\downarrow\downarrow\uparrow\downarrow$, with energies $J^{(1)}_1-J^{(1)}_2-J^{(2)}_1+3(-J^{(2)}_2-J^{(3)}_1+J^{(3)}_2)-h$ and its cyclic permutations.

\section{Ground states}
Ground states are obtained from irreducible blocks by direct comparison of their energies in the parameter space. In the absence of magnetic field, there is a symmetry under the interchange up$\leftrightarrow$down, but in the presence of a magnetic field, the blocks with more than half of the spins pointing down cannot be ground states. In certain regions of the parameter space, which can be points or lines, there is degeneracy, i.e. there are ground states with the same energy. In these cases other ground states could be formed as follows \cite{morita}. If some of these blocks, or cyclic permutations of them, ${\cal B}_1,\dots,{\cal B}_\nu$, are such that the first $r$ spins of each of them coincide, then they can be merged into an irregular chain in any order ${\cal C}=\cdots{\cal B}_i\cdots {\cal B}_j\cdots$ and the total energy of the chain will be the sum of the energies of the blocks $E({\cal C})=\sum_i^\nu n_i E({\cal B})$, where $E({\cal B})$ is the energy of the blocks and $n_i$ their frequencies. Otherwise, if the blocks cannot be merged, the regular chains formed by these blocks coexist, i.e. they can transform into each other spontaneously. 

In the following we give the ground state diagrams for three cases, with a magnetic field $h>0$. 

\begin{figure}
\centering
\includegraphics[height=5cm,width=5cm]{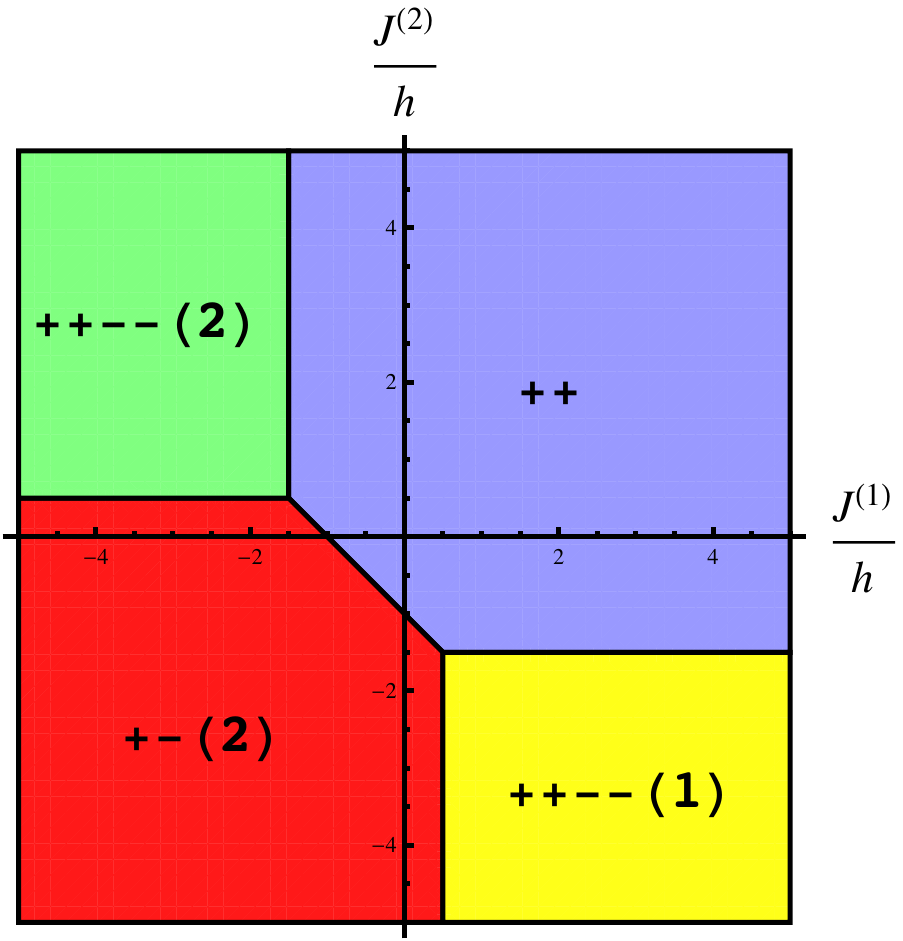}
\caption{{\protect\footnotesize {Ground states for a diatomic chain with next nearest neighbor interactions and spins $S^{(1)}=1/2$ and $S^{(2)}=1$, the first spin in the four blocks is of type (1).}}}
\label{diatomic1}%
\end{figure}

Figure \ref{diatomic1}: Diatomic chain with next nearest neighbor interactions and different spins $S^{(1)}=\pm 1$, $S^{(2)}=0, \pm 1$. The ground states are 
${\cal C}_{++}=\uparrow\uparrow$ with energy $\epsilon=-1/2(J^{(1)}+J^{(2)})-3/2\,h$, ${\cal C}_{+-}=\uparrow\downarrow$ with first spin $(2)$, with energy $\epsilon^{(2)}=1/2(J^{(1)}+J^{(2)})-1/2\,h$ and
${\cal C}_{++--}=\uparrow\uparrow\downarrow\downarrow$ with energies $\epsilon^{(1)}=-J^{(1)}+J^{(2)}$ and $\epsilon^{(2)}=J^{(1)}-J^{(2)}$

\begin{figure}
\centering
\includegraphics[height=5cm,width=5cm]{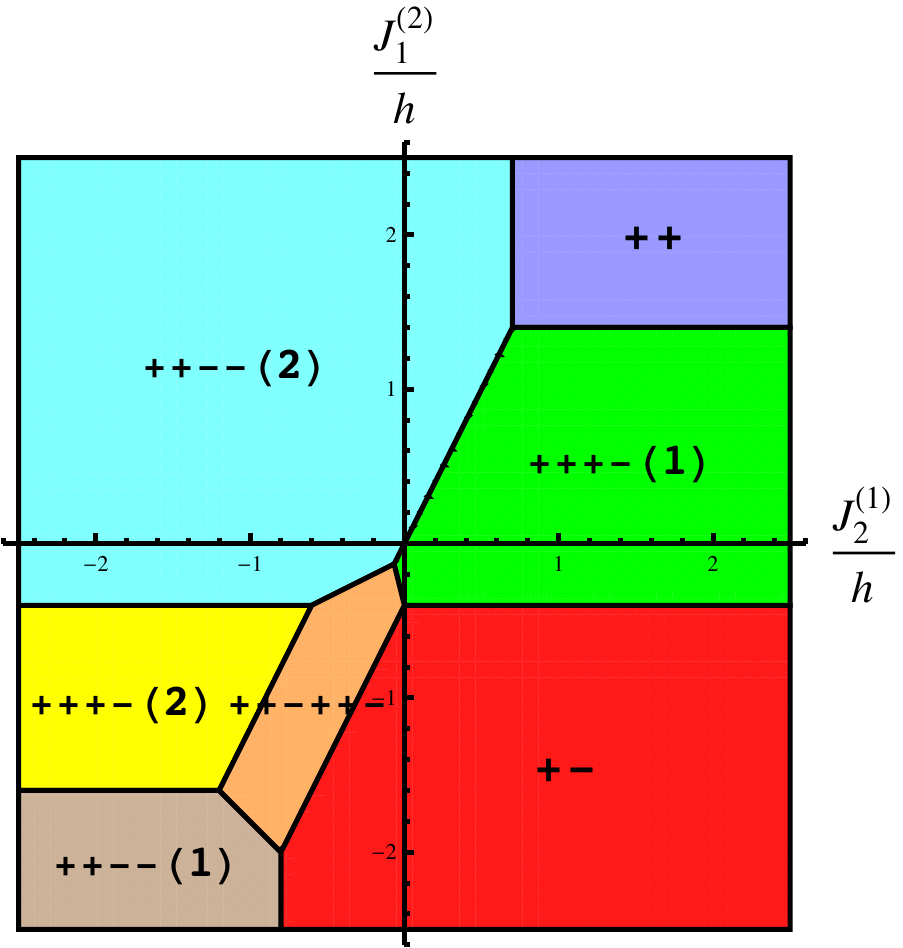}
\caption{{\protect\footnotesize {Ground states for a diatomic chain with second neighbor interactions and equal spins $S^{(1)}=S^{(2)}=\pm 1$.}}}
\label{diatomic2}%
\end{figure}

Figure \ref{diatomic2}: Diatomic chain with second neighbor interactions and equal spins $S^{(1)}=S^{(2)}=\pm 1$. The ground states are 
${\cal C}_{++}=\uparrow\uparrow$ with energy $\epsilon_{++}=-1/2(J^{(1)}_1+J^{(2)}_1+J^{(1)}_2+J^{(2)}_2)- h$,
${\cal C}_{+-}=\uparrow\downarrow$ with energy $\epsilon_{+-}=1/2(J^{(1)}_1+J^{(2)}_1-J^{(1)}_2-J^{(2)}_2)$, 
${\cal C}_{+++-}=\uparrow\uparrow\uparrow\downarrow$ with energies $\epsilon^{(1)}_{+++-}=1/2(-J^{(1)}_2+J^{(2)}_2)- h/2$ and $\epsilon^{(2)}_{+++-}=1/2(J^{(1)}_2-J^{(2)}_2)-h/2$,
${\cal C}_{++--}=\uparrow\uparrow\downarrow\downarrow$ with energies $\epsilon^{(1)}_{++--}=1/2(-J^{(1)}_1+J^{(2)}_1+J^{(1)}_2+J^{(2)}_2)$ and $\epsilon^{(2)}_{++--}=1/2(J^{(1)}_1-J^{(2)}_1+J^{(1)}_2+J^{(2)}_2)$ and  
${\cal C}_{++-++-}=\uparrow\uparrow\downarrow\uparrow\uparrow\downarrow$ with energy $\epsilon_{++-++-}=J^{(1)}_1+J^{(2)}_1+J^{(1)}_2+J^{(2)}_2-2h$.

\begin{figure}
\centering
\includegraphics[height=5cm,width=5cm]{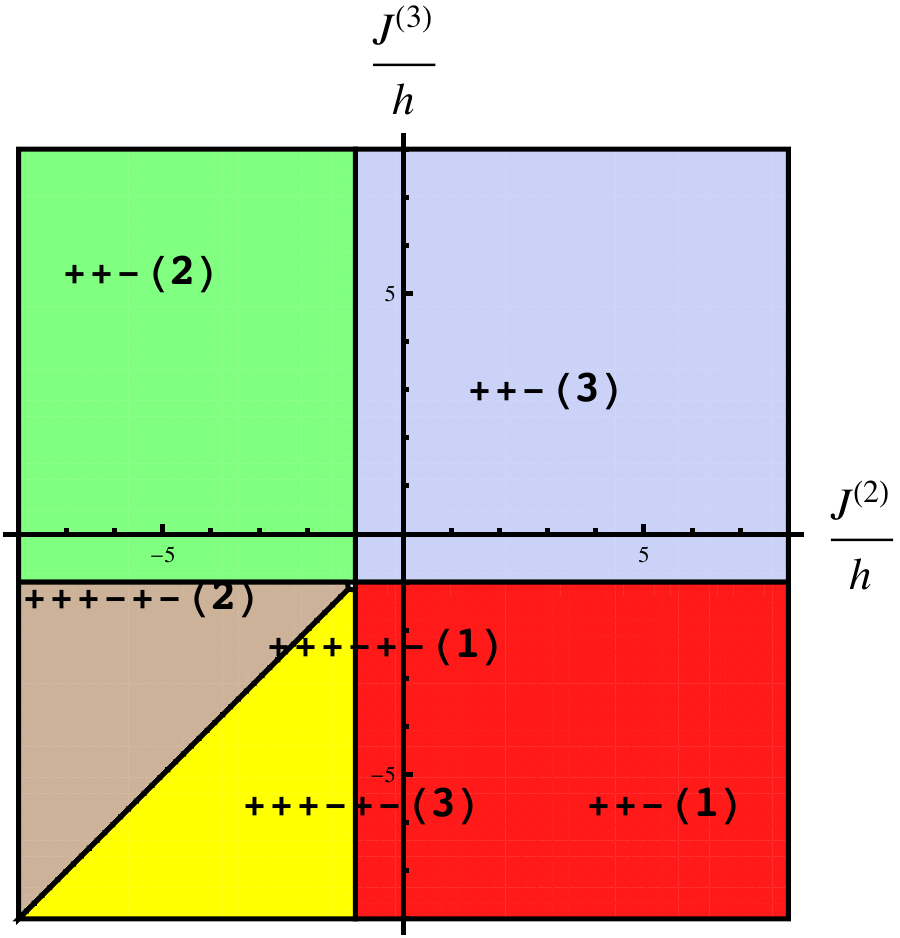}
\caption{{\protect\footnotesize {Ground states for a triatomic chain with nearest neighbor interactions and all spins $1/2$, $J^{(1)}=1$.}}}
\label{triatomic1}%
\end{figure}

\begin{figure}
\centering
\includegraphics[height=5cm,width=5cm]{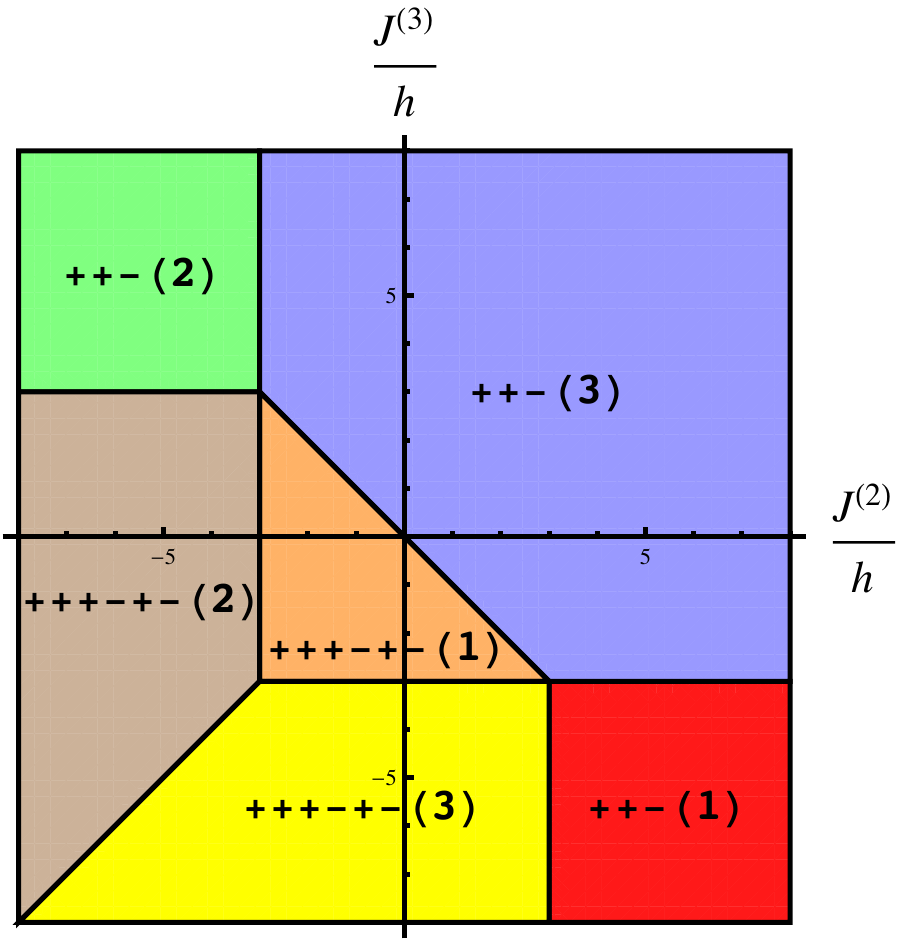}
\caption{{\protect\footnotesize {Ground states for a triatomic chain with nearest neighbor interactions and all spins $1/2$, $J^{(1)}=-3$.}}}
\label{triatomic2}%
\end{figure}

Figures \ref{triatomic1} and \ref{triatomic2}: Triatomic chain with next nearest neighbor interaction and equal spins, $S^{(1)}=S^{(2)}=S^{(3)}=\pm 1$. The ground states are ${\cal C}_{+++}=\uparrow\uparrow\uparrow$ with energy $\epsilon=-1/3(J^{(1)}+J^{(2)}+J^{(3)})-h$, 
${\cal C}_{++-}=\uparrow\uparrow\downarrow$ with energies $\epsilon^{(1)}=1/3(-J^{(1)}-J^{(2)}+J^{(3)}-h)$, $\epsilon^{(2)}=1/3(-J^{(2)}-J^{(3)}+J^{(1)}-h)$ and $\epsilon^{(3)}=1/3(-J^{(3)}-J^{(1)}+J^{(2)}-h)$ and ${\cal C}_{+++-+-}=\uparrow\uparrow\uparrow\downarrow\uparrow\downarrow$ with energies $\epsilon^{(1)}=1/3(J^{(3)}-h)$, $\epsilon^{(2)}=1/3(J^{(1)}-h)$ and $\epsilon^{(3)}=1/3(J^{(2)}-h)$. In order to visualize all the phases, we give the cases $J^{(1)}=1$, Fig. \ref{triatomic1} and $J^{(1)}=-3$, Fig. \ref{triatomic2}.

\section{Conclusions}
We have given, for polyatomic chains, a generalization of Morita's work \cite{morita} on the evaluation of a chain to obtain its ground states, based on the concept of irreducible blocks. We first review Morita's works \cite{morita,morita2}, and give an explicit demonstration of the results of the second work. By polyatomic chains we mean chains with a periodic structure, with different spins and exchange interactions of any range. The main assumption is translational invariance of the lattice, up to possible boundary terms. Moreover, the account of the interactions to the right of a spin (it could be to the left) allows for a systematic treatment. We consider first diatomic chains in several cases, and then we give the generalization to the polyatomic case, with interactions of any range. We show that, for periodicity $l$ and interactions of range $r$, the length of the irreducible blocks is a multiple of $l$ and has a maximum value $kl$, where $k$ is the maximum integer with $kl\leq l(C_{M}-1)+r$, as given at the end of Sections III and IV. The ground states must be chosen from the irreducible blocks by direct comparison of their energies. There may be degenerated states in such a way that nonregular ground state chains could be possible. In general there is no invariance under parity. For very long chains the boundary effects can be neglected, but for short chains their contribution can be important, and should be considered individually \cite{gosh}. We illustrate the results and give the ground state diagrams for diatomic and triatomic cases, with equal and different spins. These results could be applied to the study of polymers \cite{strobl} and lattice gases \cite{lee,mcquistan}. 
An interesting question would be to see if the different interactions compete and lead to frustration \cite{sandvik,kaplan}. 
The Ising model on two-dimensional stripes can be reduced to the study of chains with second nearest interactions \cite{dublenych}. It would be interesting to consider cases with different types of atoms.

\vskip 1truecm
\centerline{\bf Acknowledgements}
We thank VIEP-BUAP and PIFI-SEP for the support, P. M. thanks O. Munive and S. Bernardino for fruitful discussions.


\end{document}